# Deep-potential enabled multiscale simulation of gallium nitride devices on boron arsenide cooling substrates

Jing Wu[1†], E Zhou[1], An Huang[1], Hongbin Zhang[2], Ming Hu[3], and Guangzhao Qin[1,4,5,*]

[1] *State Key Laboratory of Advanced Design and Manufacturing Technology for Vehicle, College of Mechanical and Vehicle Engineering, Hunan University, Changsha 410082, P. R. China*

[2]*Institut für Materialwissenschaft, Technische Universität Darmstadt, Darmstadt, 64289, Germany*

[3]*Department of Mechanical Engineering, University of South Carolina, Columbia, SC 29208, USA*

[4]*Research Institute of Hunan University in Chongqing, Chongqing 401133, China*

[5]*Greater Bay Area Institute for Innovation, Hunan University, Guangzhou 511300, Guangdong Province, China*

[†] *Present address: School of Energy and Power Engineering, Huazhong University of Science and Technology, Wuhan, Hubei 430074, China*

**Abstract:** High-efficient heat dissipation plays critical role for high-power-density electronics. Experimental synthesis of ultrahigh thermal conductivity boron arsenide (BAs, 1300 W $m^{-1}K^{-1}$) cooling substrates into the wide-bandgap semiconductor of gallium nitride (GaN) devices has been realized. However, the lack of systematic analysis on the heat transfer across the GaN-BAs interface hampers the practical applications. In this study, by constructing the accurate and high-efficient machine learning interatomic potentials, we performed multiscale simulations of the GaN-BAs heterostructures. Ultrahigh interfacial thermal conductance (ITC) of 260 MW $m^{-2}K^{-1}$ is achieved, which lies in the well-matched lattice vibrations of BAs and GaN. The strong temperature dependence of ITC was found between 300 to 450 K. Moreover, the competition between grain size and boundary resistance was revealed with size increasing from 1 nm to 1000 μm. Such deep-potential equipped multiscale simulations not only promote the practical applications of BAs cooling substrates in electronics, but also offer new approach for designing advanced thermal management systems.

[*] Author to whom all correspondence should be addressed. E-Mail: gzqin@hnu.edu.cn

## Introduction

Thermal management is critical for electronic devices[1–7], such as high-electron-mobility transistors (HEMTs), high-power density field-effect transistors (FETs), and radiofrequency (RF) devices, where high temperature degrades the performance of the system. However, in modern semiconductor industries[8], the high hot-spot temperature in growing miniaturized electronic devices leads to a significant challenge for heat dissipation[9,10] as shown in Fig. 1(a). Thus, there is a strong impetus to search for high thermal conductivity (HTC) substrate materials, which can be integrated with hot-spot units for efficient heat dissipation. Therein the emerging interfacial thermal resistance (ITR) at the interface of heterostructure[11–16] plays a crucial role, which may hamper the heat transfer. From micro-/nanoscale perspective, some phonons carrying heat fail to cross the interface, resulting in the emerging interfacial thermal resistance as illustrated in Fig. 1(d) and the significant temperature jump as shown in Fig. 1(e). Thus, both HTC and low ITR are essential to heat dissipation.

Significant efforts have been employed to search, design, and investigate HTC materials for efficiently cooling down electronic systems[17–28]. Recently, some HTC compounds have been predicted based on first principles calculations. For instance, binary boron compounds and ternary boron compounds are reported[17,29,30], both with thermal conductivity ($\kappa$) higher than 500 W $m^{-1}K^{-1}$. Especially, a remarkably high $\kappa$ of 1300 W $m^{-1}K^{-1}$ for boron arsenide (BAs) has been confirmed by both theoretical calculations[31,32] and experimental measurements[25–27]. Recently, experimental studies show the high carrier mobility of BAs[33,34], which indicates promising applications of BAs in high-power electronics and optoelectronics. Commonly, the high-power density electronic devices of HEMTs are composed of gallium nitride (GaN), silicon, and silicon carbide substrates, where the performance bottleneck generally exists due to the low thermal conductivity. Recently, Kang. *et al.*[35] successfully fabricated GaN devices on BAs cooling substrates in experiments. The crystal structures of BAs and GaN are presented in Figs. 1(b) and (c), respectively. Outstanding cooling efficiency compared with diamond-supported devices was observed, which is benefited from the low ITR in addition to the ultrahigh $\kappa$ of BAs. In fact, boundaries and interfaces are inevitable for the practical applications of cooling substrates in microelectronics. The interfacial thermal transport across GaN and the neighboring BAs is important for a high-performance thermal management, especially for the near-junction region. However, limited by the accurate interatomic interaction potential, the lack of systematic analysis on the heat transfer across the GaN-BAs interface hampers the practical applications. The fundamental understanding on the heat transfer across the GaN-BAs interface can offer insight for the rational design of advanced thermal management systems, which demands for a systematic study of the ITR. We have pioneered to construct an innovative multiscale simulation framework that combines DFT calculation, Molecular dynamics (MD) simulation, Monte Carlo (MC) simulation, and FE simulation, with the aid of the *state-of-the-art* machine learning technique. The simulation framework enables deep understanding of the characteristics and principles of materials at different spatial and time scales.

MD simulations have been widely employed to study the thermal transport properties of interfaces in heterostructures[11,36,37]. Although it is robust and "automatic" to consider the effect of surface and interface, one apparent disadvantage of MD simulations is that it largely relies on an accurate interatomic potential. Unfortunately, the interatomic potentials of BAs are lacking up to now, which limits a detailed theoretical analysis on the interfacial thermal transport properties and thus hampers the potential applications of BAs as a competing cooling substrate for microelectronics. Previous studies used diffuse mismatch mode (DMM) and acoustic mismatch model (AMM) to calculate the ITC. Both methods are empirical models that only consider the basic phonon scattering property of the two in-contact materials and assume the interface is perfect, while the detailed atomic structure at the interface does not play any role in the models. Compared to the traditional methods, direct MD simulations based on accurate machine learning (ML) potential are capable of studying the phonon thermal transport across the GaN-BAs interface with all order anharmonic phonon scattering involved and the simulation are much closer to the physical interface. Recently, it has been demonstrated that ML techniques can be utilized to construct accurate interatomic potentials[38–40], where artificial neural networks (ANN) and neural network potentials (NNPs) are the most widely used methods[41,42]. Correspondingly, the resulting multiscale simulations based on NNPs combine the advantages of the accuracy of *ab-initio* calculations and the efficiency of analytically parameterized potentials. Thus, by empowering the multiscale study of the thermal transport properties of heterostructure with first principles accuracy[43,44], the NNPs land the foundation to promote in-depth understanding and predictive design of efficient heat dissipation for engineering microelectronic devices.

In this study, we constructed the NNPs for GaN, BAs and their interfaces to systematically investigate the thermal transport properties of GaN-BAs heterostructure. The bulk thermal transport properties of GaN and BAs were firstly evaluated and validated with the existing reports. We then employed non-equilibrium MD (NEMD) and MC simulations to evaluate and analyze the interfacial thermal transport properties of the GaN-BAs heterostructure. The NEMD results reveal significant temperature-dependent interfacial thermal conductance between 300 and 450 K in BAs-GaN, suggesting enhanced phonon transmission and inelastic phonon scattering. Finite element models (FEM) were further constructed to simulate GaN-BAs heterostructure, which is similar to the amorphous or polycrystalline layer near interface. Finally, we report the competition between grain size and boundary resistance in our polymer model expanding the size-scale from 1 nm to 1000 μm. The work is expected to shed light on practical thermal management engineering with interfaces.

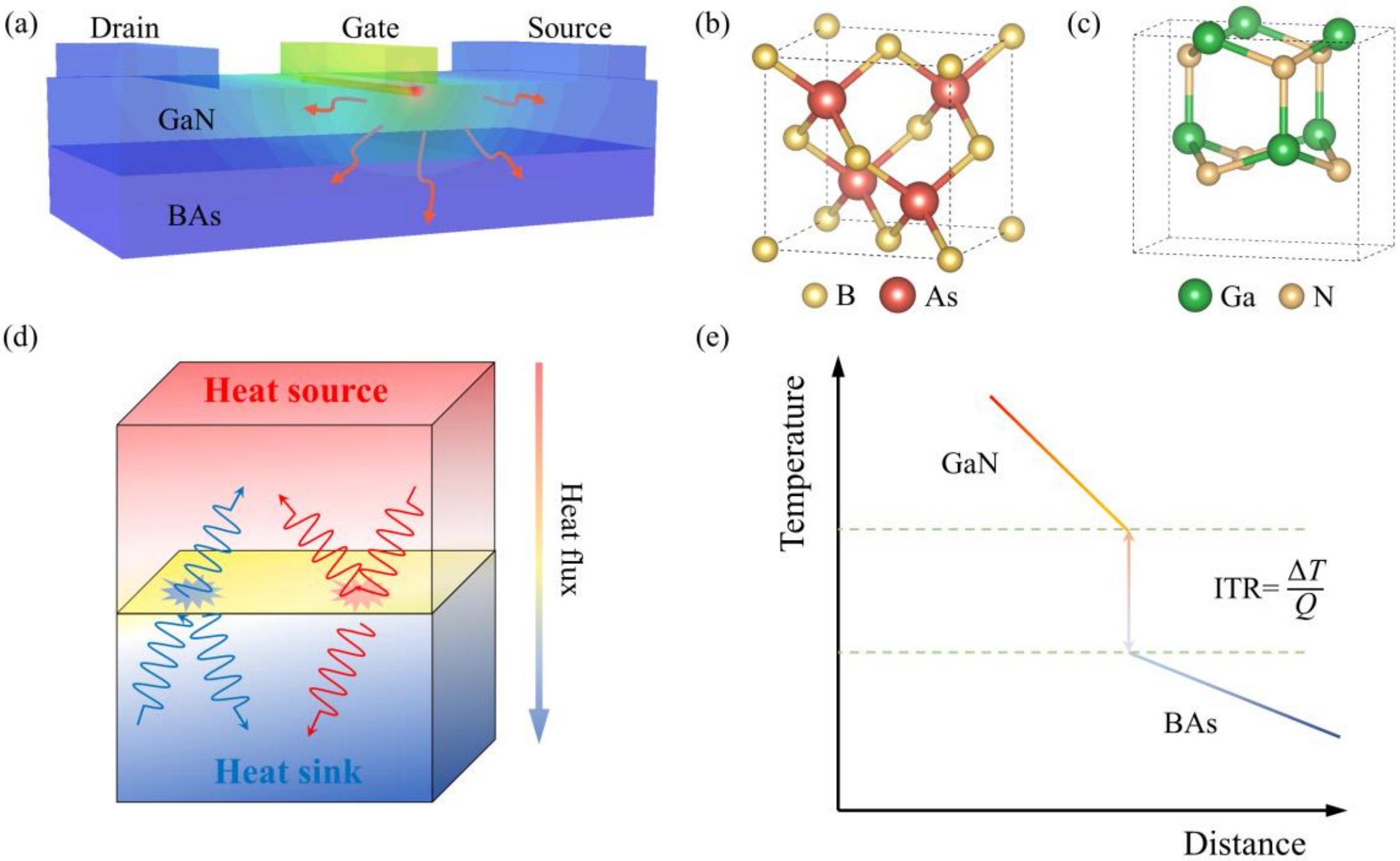

**Fig. 1** Thermal management employing the boron arsenide (BAs) cooling substrates for gallium nitride (GaN) devices. (a) The schematic of BAs supporting GaN HEMTs and the distribution of hot-spot temperature. (b-c) The crystal structures of unit cell of BAs and GaN, respectively.

(d) The phonon transmission and reflection models at the heterostructure interface. (e) The steady state of temperature profile for GaN-BAs heterostructure and the formula of interfacial thermal resistance (ITR), where $Q$ and $\Delta T$ are the heat flux and temperature drop, respectively.

## Results and discussions

### Neural network potential (NNP) training

To investigate the heat transfer across the GaN-BAs interface as shown in Fig. 1, we firstly construct the NNPs for BAs with 128 atoms, GaN with 144 atoms, and the GaN-BAs heterostructure with 128 atoms. For the atoms at the interface, we visualized the energy, force, and stress distribution as shown in Fig. S4, Fig. S5, and Fig. S6, respectively. The selection of configurations used in neural network (NN) training process is essential to construct reliable NNP models. More importantly, the interfacial structures in the GaN-BAs heterostructures can be complex. Therefore, the active learning strategy is utilized during the construction of NNPs, which shows high efficiency and low computation costs to collect relevant training configurations for pristine BAs, GaN, and GaN-BAs heterostructures. As shown in Fig. 2(a), the NNP training workflow comprises three steps (see the section of Methods for more details). With the increasing iterations of the *ab initio* calculations, we are able to collect diverse snapshots for the NNPs training. Finally, 2250, 2535, and 10350 training snapshots were collected for pristine BAs, GaN, and the GaN-BAs heterostructures, respectively.

The quality of the resulting NNP models is verified by performing a comparison for energies and forces ($f_x$, $f_y$, $f_z$) between density functional theory (DFT) calculations and NNPs predictions. The test configurations are composed of 2103, 2022 and 1990 snapshots for BAs, GaN and the GaN-BAs heterostructures, respectively, which are not included in the configurations for previous training. From Figs. 2(d) and 2(i), it is obvious that NNPs are able to accurately reproduce the energies and forces obtained from DFT calculations. The force root mean square error (RMSE) are 18, 22, and 36 meV/Å for BAs, GaN, and GaN-BAs heterostructures, respectively. The small RMSE also implies the high quality of the NNP models. On the other hand, despite the successful description of the interface of the trained model, we have to say that the model is trained with limited data, and it surely fails in certain cases due to the native weakness of transferability from machine learning technologies, such as interfacial microstructures, doping elements, and disorder structures. With the NNP models, we calculate the phonon dispersions of BAs and GaN, and the results coincide with those from explicit DFT calculations, as comparably shown in Figs. 2(b) and 2(c). Moreover, both BAs and GaN show obvious phonon gap, especially for BAs, which implies strong high order phonon scattering[31].

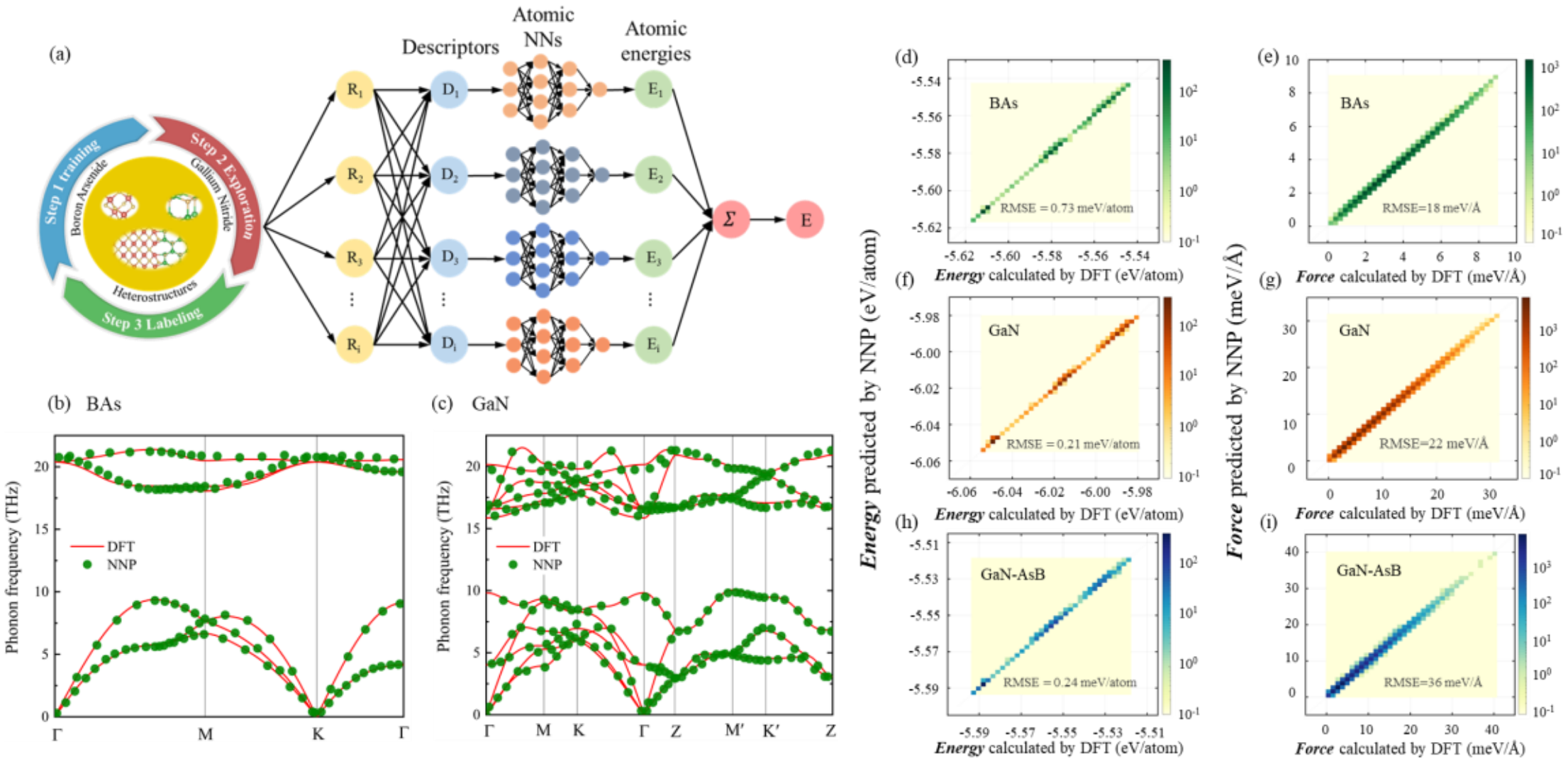

**Fig. 2** The active learning workflow and the accuracy verification of the constructed NNP. (a) The integration of three steps of the training of NNP models, exploration of new configuration, and calculation of energies and forces at DFT level. Besides, the structure of the NNP models is also presented. (b-c) The comparison of phonon dispersion of (b) BAs and (c) GaN between NNP and DFT. (d-i) The comparison of energies and forces of BAs, GaN, GaN-BAs heterostructures from NNP predictions and DFT calculations, where color indicates the number density of points.

### Thermal transport in GaN and BAs

With the well-trained NNP models, we can evaluate the lattice thermal conductivities ($\kappa$) of GaN and BAs. The finite displacement method was adopted to calculate the second- and third-order constants that are required to solve the Boltzmann transport equation (BTE)[46–48]. And the force calculations are performed using the Large-scale Atomic/Molecular Massively Parallel Simulator (LAMMPS) software package[49] coupled with our trained NNPs. Fig. 3(a) shows the anisotropic $\kappa$ of GaN considering four-phonon scattering and isotope effects, where the $\kappa$ along the $a$ and $c$ axis are 217 and 229 W $m^{-1}K^{-1}$, respectively. These results are in excellent agreement with previously reported DFT calculations[50] ($a$ = 223 and $c$ = 243 W m$^{-1}$K$^{-1}$) and experimental measurements[51-53]($a$ = 217 and $c$ = 228, 195 W m$^{-1}$K$^{-1}$). Recent experimental measurements and theoretical calculations demonstrated that BAs has strong four-phonon scatterings[25,27,54]. Consequently, after considering four-phonon scatterings, the $\kappa$ of BAs is calculated to be 1510 W m$^{-1}$K$^{-1}$ using NNP model of BAs at room temperature [Fig. 3(b)], which is consistent with the reported value (1400 W m$^{-1}$K$^{-1}$) in literature[31]. Note that we can obtain the calculation of the fourth force constants in a few minutes when using NNP model to evaluate atomic forces in supercells, which achieves a significant reduction of computational cost with similar precision compared to full first-principles calculations. The MD simulations of interfacial thermal transport across the GaN-BAs interface naturally includes all orders phonon transport, which is much closer to the physical interface compared to other methods. Different from the complex procedure of including four-phonon scatterings in BTE[55], it is natural for MD simulations to consider the four-phonon scatterings and even higher order scattering.

To obtain the accurate $\kappa$ of BAs from classical MD simulations, we performed the equilibrium MD (EMD) simulations based on the Green-Kubo (GK) method using the NNP models, where the thermal conductivity is evaluated based on the heat current autocorrelation function (HCACF):

$$\kappa = \frac{V}{3k_B T^2}\int_0^\infty \left\langle J\left(0\right)\cdot J\left(t\right)\right\rangle dt\,, \tag{1}$$

where $V$ is the volume, $k_B$ is the Boltzmann constant, $T$ is the temperature, $J$ is the heat flux, and $\langle\ \rangle$ represents the time average. The resulting HCACF is displayed in Fig. 3(c). It is shown that 1200 ps EMD simulations can enable the generation of a converged $\kappa$ of BAs. For obtaining the HCACF $\langle J(0)\cdot J(t)\rangle$, the heat flux $J$ is calculated as follows:

$$J = \frac{1}{V}\sum\nolimits_i (E_i \boldsymbol{v}_i - \boldsymbol{S}_i\cdot\boldsymbol{v}_i)\,, \tag{2}$$

where $E_i$ and $\boldsymbol{v}_i$ are the atomic potential energy and velocity, respectively. The atomic virial stress tensor $S_i$ is defined as the outer product of relative atomic position $\boldsymbol{r}_i - \boldsymbol{r}_j$ and the derivative of local potential energy with respect to the neighboring atom position:

$$\boldsymbol{S}_i = \sum_i (\boldsymbol{r}_i - \boldsymbol{r}_j) \otimes \frac{\partial E_i}{\partial (\boldsymbol{r}_j - \boldsymbol{r}_i)} \quad . \tag{3}$$

In the MD simulations, each particle is initially assigned a random velocity using the velocity random seeds. The ensemble of the generated velocities is a uniform distribution, which can be scaled to produce the requested temperature. The introduction of velocity random seeds allows the system to evolve from a uniformly distributed state rather than from a specific ordered state. Such initial conditions help simulate the behavior of the system under different conditions and explore all the possible states and dynamics, as widely utilized in previous studies[56–59]. To reduce the simulation uncertainty and obtain reliable result, the $\kappa$ is obtained by averaging 14 independent simulations as shown in Fig. 3(d), where the shaded area indicates the statistical uncertainty on the mean value. As summarized in Table 1 including the results from previous calculations[31,32,60] and experiments[25–27], the $\kappa$ of BAs obtained from MD simulations in this work (1179 ± 91 W $m^{-1}K^{-1}$) is well consistent with previous reports, especially compared with the BTE results involving four-phonon scattering and the experimental measurements. The length-dependent thermal conductivity of GaN and BAs is also calculated in the Fig. S9. Note that it is for the first time the ultrahigh $\kappa$ of BAs is verified from direct MD simulations of atomic motions. Besides, we also study other phonon related properties of GaN and BAs, such as group velocity, phonon relaxation time, and the contribution to $\kappa$ from different phonon branches, which are shown in the Fig. S1 and Fig. S2. Especially, the $4^{th}$ phonon scattering phase space for BAs is also provided for further understanding.

**Table. 1** The comparison of $\kappa$ (BAs) from MD simulations with previous reports from DFT calculations and experimental measurements. The 3-ph and 4-ph represent the phonon scattering involving three- and four-phonons, respectively.

| | Calculation | | | | Experiment | | |
|---|---|---|---|---|---|---|---|
| | this work | Liu. *et al*[32] | Yang. *et al*[60] | Feng. *et al*[31] | Li. *et al*[27] | Tian. *et al*[25] | Kang. *et al*[26] |
| $\kappa$ (W m$^{-1}$k$^{-1}$) | 1179 ± 91 (MD) | 2381 (3-ph)<br>1182 (3-ph+4-ph) | 2276 (3-ph)<br>1441 (3-ph+4-ph) | 2241 (3-ph)<br>1417 (3-ph+4-ph) | 1000±90 | 1160±130 | ~1300 |

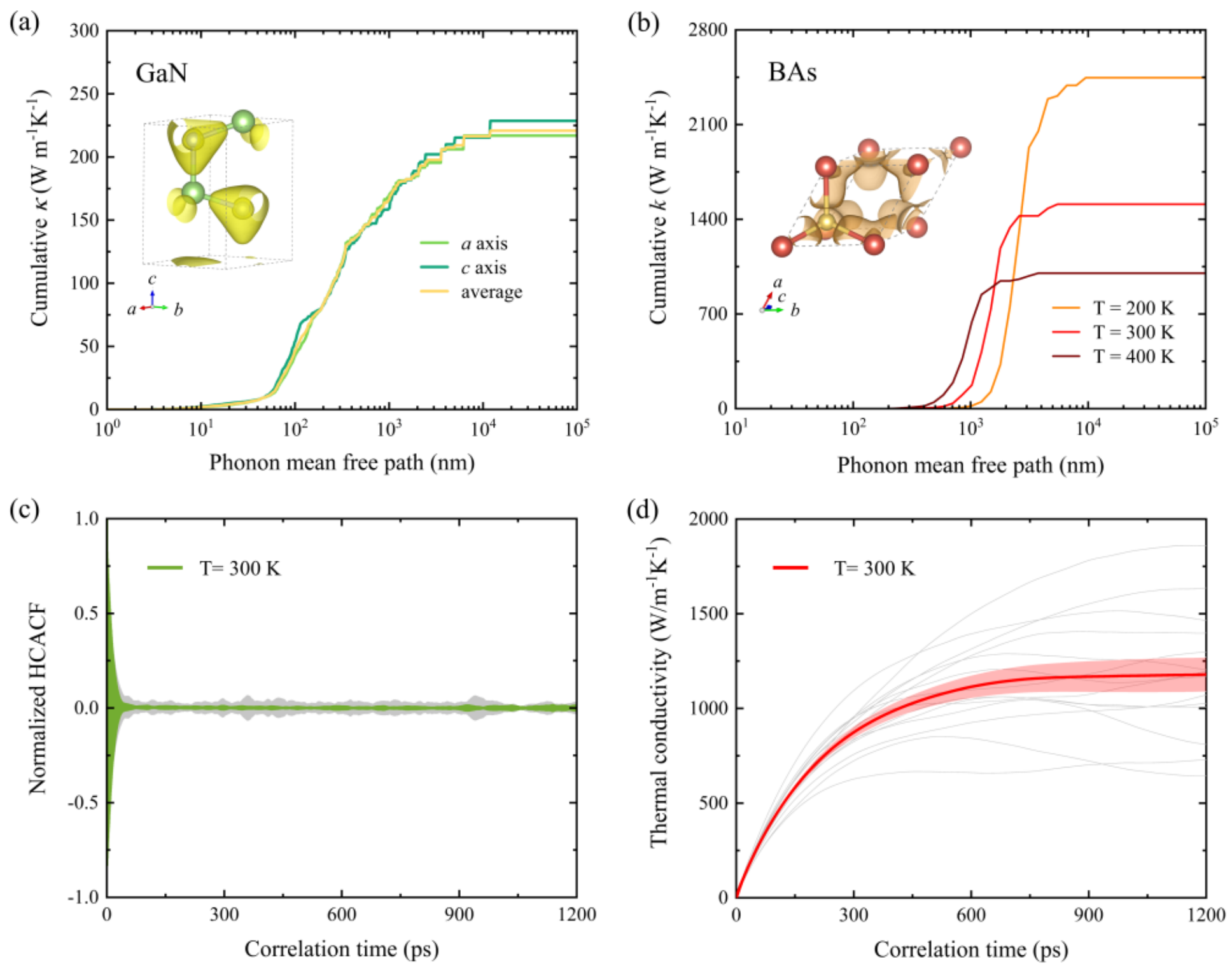

**Fig. 3** The thermal conductivity ($\kappa$) of GaN and BAs. (a-b) Comparison of the accumulative $\kappa$ with respect to phonon mean free path (MFP) for (a) GaN with anisotropy and (b) BAs at different temperatures, respectively. The insets are electron localization function (ELF). (c-d) The converging (c) normalized HCACF and the (d) corresponding $\kappa$ at 300 K for BAs as a function of simulation time, where the shaded area indicates the statistical uncertainty on the mean value from 14 independent simulations.

**The interfacial thermal conductance (ITC) of the GaN-BAs heterostructures**

The interfacial heat transfer in the GaN-BAs heterostructures is further investigated by performing NEMD simulations. We put GaN on heat source and BAs on heat sink, as shown in Fig. 4(a), which shows a stable temperature profile of a steady state. There is an obvious temperature jump ($\Delta T$) at the interfaces, implying a finite ITR between BAs and GaN. Thus, the ITC of the heterostructures is evaluated as 260 MW $m^{-2}K^{-1}$ at 300 K, which is in good agreement with the result[35] measured by time-domain thermoreflectance (TDTR) (250 MW $m^{-2}K^{-1}$) as collected in Fig. 4(b). The relatively higher ITC in calculations may be due to the fact that there exists epilayer between GaN and BAs in the experimental measured samples[35]. It is worth pointing out that the ITC result may be affected by size due to the large phonon mean free path of BAs, and the size of the model used in this study is 30 nm (the length-dependent ITC is presented in Fig. S8), which is similar as the heterostructure between graphene and boron nitride[36]. We also perform the calculations with GaN on heat sink and BAs on heat source, and the results are consistent, which indicates weak thermal rectification effect in the GaN-BAs interface.

It can be seen that ITC gradually increases with respect to the increasing temperature ranging from 300 to 450 K. The strong temperature dependence of ITC is different from the most interfaces including GaN-diamond and GaN-SiC, where the ITC is almost independent of temperature in the application temperature range[10, 45]. To explore the underlying mechanisms of the temperature dependent ITC in BAs-GaN, the spectral heat flux at different temperatures is plotted in Fig. S7(a). With the increasing temperature, spectral heat flux gradually increases both at low and high frequencies and then converges, which are similar to the variation of ITC as shown in Fig. 4(b). The phenomenon means that more heat carries are activated to across the GaN-BAs interface by the increased temperature, inducing more inelastic phonon interface scattering, and thus enhances ITC by enhancing phonon transmission coefficients [Fig. S7(b)].

To fundamentally understand the behavior of ITR, the phonon eigenvectors of the GaN-BAs heterostructure interface at different frequencies are visualized in Fig. S3. There are enormous phonon eigenvectors for both GaN and BAs, which indicate the active phonon vibration figure for 5 THz. In other frequencies, phonon eigenvectors are nearly negligible. It is worth noting that the phonon eigenvectors of GaN is strong at 20 THz, which is consistent with the large phonon relaxation time as shown in Fig. S1(b). Besides, the phonon density of states (PDOS) was explored for GaN and BAs. As shown in Fig. 4(c), there is an obvious band gap ranging from 10 to 16 THz, coincided with the phonon dispersions in Figs. 2(b) and 2(c). Moreover, the large overlap area of PDOS indicates the well-matched lattice vibrations of GaN and BAs. Thus, the small ITR and the corresponding high ITC are achieved in the GaN-BAs heterostructures. Such conclusions would benefit the potential applications of BAs as a competing cooling substrate to diamond with the advantage of not needing epilayer[7,61].

To further figure out the heat flux dependence on frequency, we use MC simulations to calculate the distribution of heat flux across the heterostructure interfaces, and the results are shown in Fig. 4(d). The frequency resolved heat flux from GaN to BAs is scaled by the thermal conductivity of materials. Apparently, there is a wide distribution range of heat flux for BAs, and the heat flux of GaN concentrates at low frequencies. Most importantly, the heat flux in the low frequency exhibits an excellent match across the interface, which explains the high ITC in the GaN-BAs heterostructures. Such behavior also implies that the low frequency phonons make greatly contributions to the ITC.

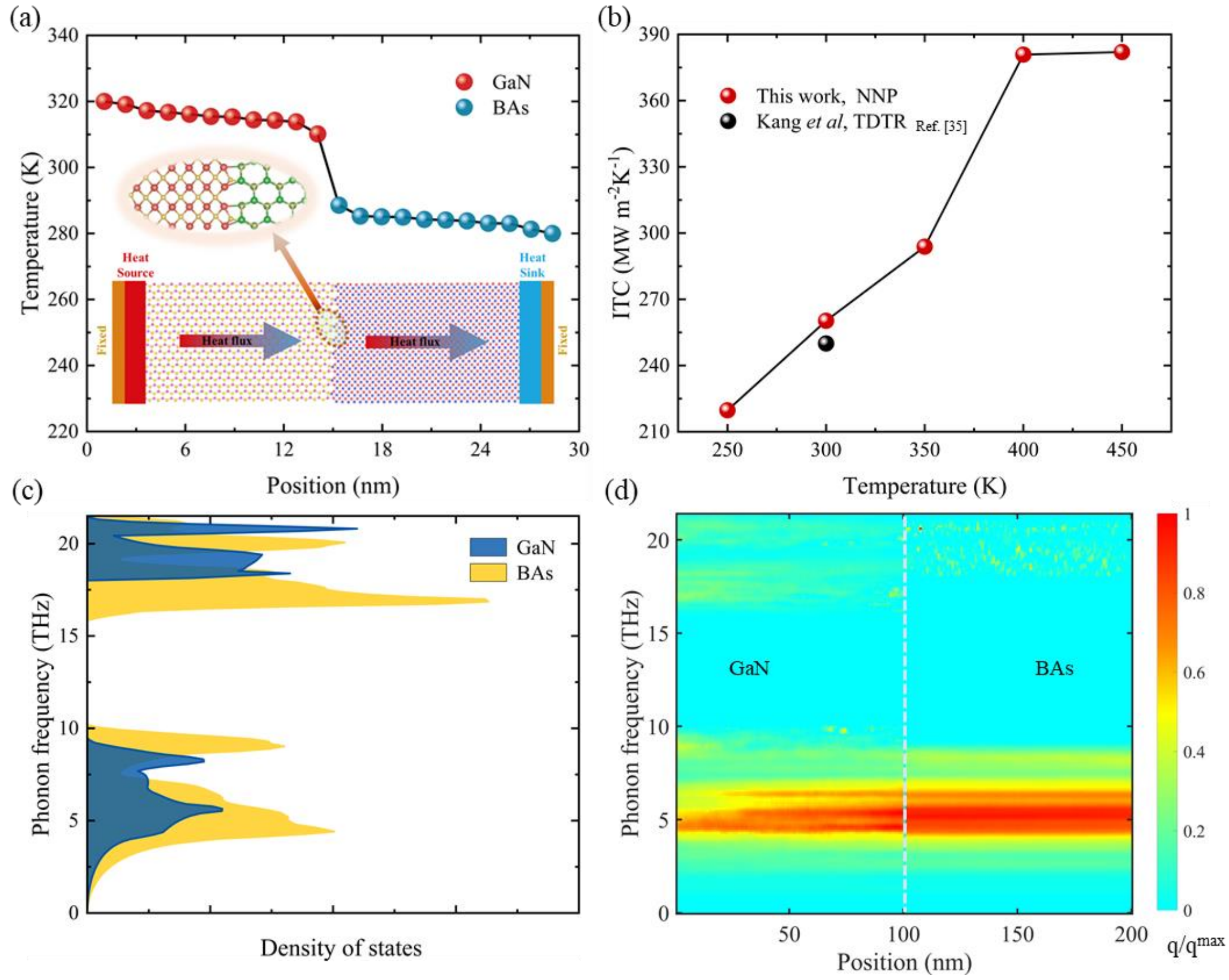

**Fig. 4** The analysis of thermal transport in GaN-BAs heterostructure from MD and MC simulations. (a) The temperature distribution in the GaN-BAs heterostructure in steady state. Inset is the model for NEMD simulations. (b) The ITC with respect to temperature, in comparison with experimental measurements. (c) The comparison of PDOS between BAs and GaN. (d) The spectral heat flux in GaN-BAs heterostructures, where the dash line indicates the interface and the color indicates the normalized density of heat flux.

**Multiscale modeling from 1 nm to 1000 μm**

Based on the understanding of heat transfer process achieved from above DFT and MD simulations, the FEM can be employed to study the heterostructure models at a macroscopic scale. The combination of FEM and MD simulations is a promising approach to overcome the multiscale effects of grain sizes and the huge computational cost. Compared with MD and DFT simulations limited in the nanoscale, the macroscopic FEM can cohere with the practical applications and get the effective $\kappa$ of GaN-BAs heterostructures in multiscale. Herein, a 1000 nm model is constructed with 50% BAs and 50% GaN as shown in Fig. 5(a). To calculate the effective $\kappa$ of such a heterostructure model, one dimensional Fourier law is applied as $\kappa_{eff} = q\frac{L}{\Delta T}$, where $q$ is the heat flux across the model. The corresponding steady states are guaranteed when the value of the inward flux is equal to that of the outward flux. The $\Delta T$ is the temperature drop between the top side and bottom side, and Fig. 5(b) shows the resulting temperature profile.

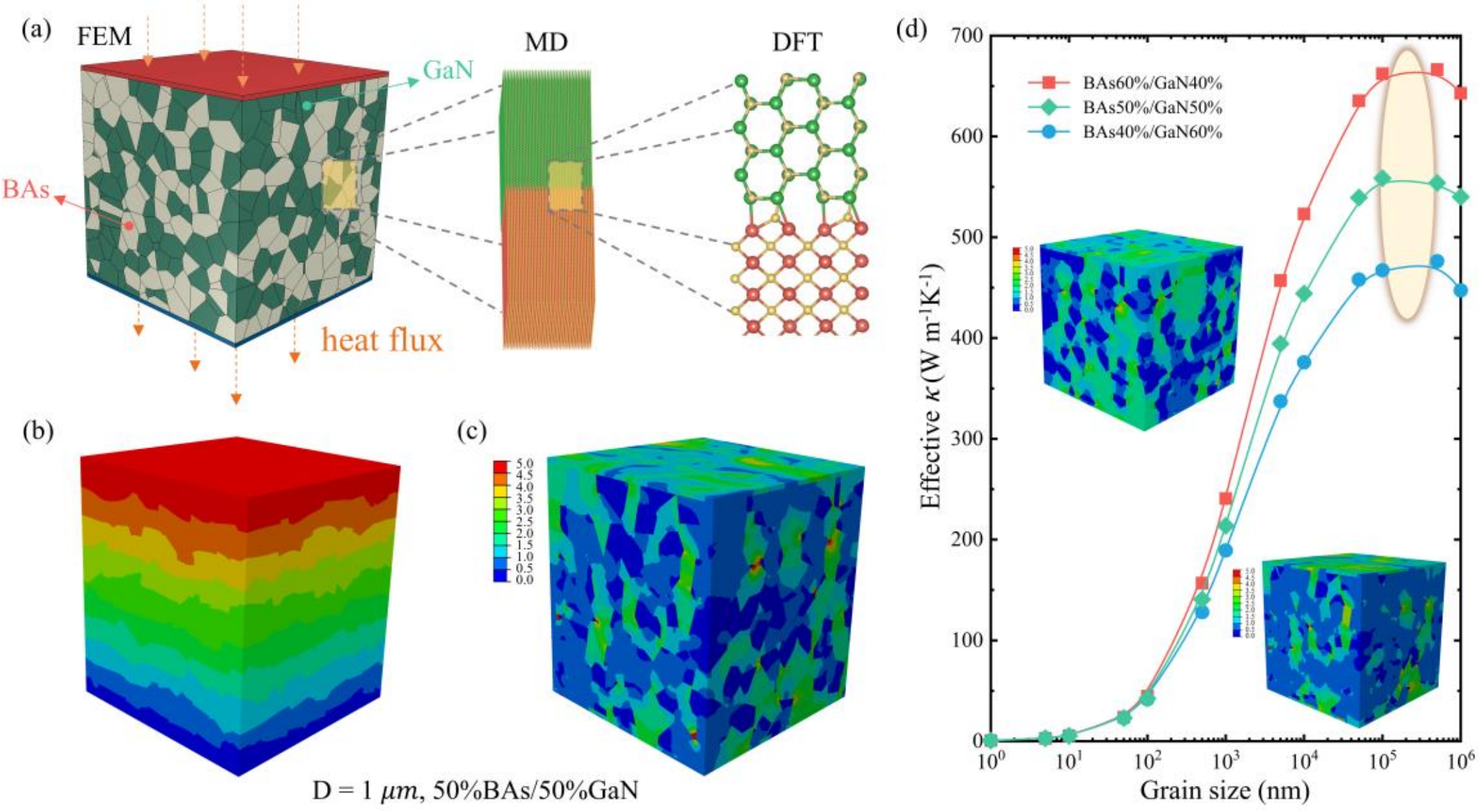

**Figure. 5** The FEM models and the effective $\kappa$ for different percentages of BAs and GaN. (a) The FEM model with 50% BAs and 50% GaN, and the multiscale Schematic. (b-c) The temperature profile and the distribution of heat flux at the typical size of 1 μm. (d) The effective $\kappa$ of different proportion BAs and GaN models with respect to the grain size. The distribution of heat flux of 40% BAs - 60% GaN, and 60% BAs - 40% GaN are shown as inset. The colored

ellipse marks the turning point of the effective $\kappa$ with the increasing grain size.

From the FEM, Fig. 5(c) shows the ununiform heat flux distribution of 50% BAs and 50% GaN, respectively. It is observed that BAs dominates the majority of heat flux, which lies in that the thermal conductivity of BAs is significantly higher than that of GaN. Moreover, similar phenomena can also be seen from the two insets in Fig. 5(d), which consist of different percentages of GaN and BAs. To further understand the mechanism of grain size dependence of the effective $\kappa$, we perform a comparison study with grain size ranging from 1 nm to 1000 μm. As shown in Fig. 5(d), there exists an important competition between the grain size and grain boundary. Firstly, when the grain size is below 10 nm, the effective $\kappa$ is insensitive to grain size. Such a behavior lies in the competition between the increasing size benefiting phonon transport and the tiny effective $\kappa$, which is due to the decreased $\kappa$ values at the much smaller size than the representative MFP as indicated in Figs. 3(a) and 3(b). Secondly, when the grain size ranges from 10 nm to 100 μm, more and more phonons are gradually activated in GaN and BAs to participate in the heat transfer with the increasing grain size. Thus, the increasing grain size leads to increased $\kappa$ in both GaN and BAs, and the role of grain boundary becomes relatively less significant, leading to the significantly increased effective $\kappa$. Thirdly, when the grain size ranges from 100 to 500 μm, the effective $\kappa$ tends to saturate, which lies in the grain size greatly exceeding the phonon MFP of GaN and BAs [Figs. 3(a) and 3(b)], and the largely weakened effect of grain boundary resistance. Finally, it is obviously seen that when the grain size over 500 μm, the effective $\kappa$ decreases with the increasing grain size. The underlying mechanism lies in that the phonon scattering dominates the thermal transport, where the increasing grain size results in more significant ITR, and the $\kappa$ of GaN and BAs hardly increases. The results as discussed above suggest that the grain size between 10 nm and 100 μm is the most favorable size scale for engineering phonon thermal transport in GaN and BAs systems.

## Conclusion

In summary, we have studied the thermal transport of GaN devices on BAs cooling substrates for high-performance thermal management. The *state-of-the-art* computational methods, *i.e.*, ML, DFT, MD, and FEM, are employed to develop systematic multiscale simulations of the thermal transport properties together with quantitative evaluations and in-depth mechanistic evaluations. The comparison of DFT and MD simulations, and the excellent agreement with the reports from previous calculations and experiments show that the trained NNP models in this study are of high accuracy. And it is *for the first time* the ultrahigh thermal conductivity of BAs is verified from direct MD simulations of atomic motions. Ultrahigh ITC of 260 MW $m^{-2}K^{-1}$ is obtained in the GaN-BAs heterostructures, which agrees well with experimental measurements, promising the conducive heat dissipation. Detailed analysis reveals that the underlying mechanism lies in the well-matched lattice vibrations of BAs and GaN. In addition, the NEMD results also show relativity large temperature dependent ITC between 300 and 450 K, which is different from previously reported interfaces, such as GaN-diamond and GaN-SiC. Based on the understanding of heat transfer process achieved from nanoscale DFT and MD simulations, we successfully established the macroscopic FEM models of the GaN-BAs heterostructures and investigated the mechanism for the grain size dependent effective $\kappa$ with the grain size ranging from 1 nm to 1000 μm. Our results provide holistic and fundamental insights into the interfacial heat transfer of the GaN-BAs heterostructures, which is expected to promote the practical applications of BAs as a competing cooling substrate to diamond for electronic devices. Most importantly, we believe that our approach based on machine learning interatomic potential driven multiscale simulations can be applied to engineer and design sophisticated thermal management systems, and hence further promote the development of miniaturized electronic devices.

# Methods

**DFT calculation**

The Vienna *ab-initio* simulation package (VASP), a widely employed *ab initio* plane wave electronic structure code for atomic scale materials modelling, was used to perform first-principles calculations with the Perdew–Burke–Ernzerhof functional of generalized gradient approximation[62]. For the DFT calculations, the kinetic energy cutoff was set as 1000 eV, a 10 × 10× 6 Monkhorst-Pack[63] *k*-point grid are employed for BAs and GaN, respectively. For self-consistent field (SCF) calculations with convergence threshold of $10^{-6}$ eV, kinetic energy cutoff was set as 600 eV, and a 1 × 1 × 1 Monkhorst-Pack *k*-point grid is employed.

**The construction of NNP models**

To collect diverse training data, active learning workflow is adopted[64]. To initialize the training data, the supercell is firstly relaxed and compressed uniformly with a scaling factor α ranging from 0.98 to 1.02, where the interval is 0.02. Then, the atomic positions and cell vectors are randomly perturbed. The magnitude of perturbations is 0.01 Å for the atomic coordinates, which is 3% of the cell vector length for the simulation cell. 15 randomly perturbed structures were created for GaN, BAs, and heterostructure, respectively. They are then utilized to perform a 10-step canonical *ab-initio* molecular dynamics simulation at T = 50 K with VASP. For the training, the smooth edition of the NNP is employed. A skip connection is adopted between two neighboring fitting layers. The cutoff is set to 7 Å. Four neural network models with Adam optimizer only combined with different parameter initializations, the embedding net with size of {30,60}, and the fitting net with size of {240, 240, 240} were trained. To the exploration step, when the force deviation of configurations is in the trust range [0.05, 0.2], it will be labeled. During the Labeling step, VASP was adopt to performance first-principles calculations. Finally, the active learning cycles are performed 28 times, 21 times, and 29 times for GaN, BAs, and GaN-BAs heterostructure, respectively. With the collected data, an accurate NNP model was trained with learning rate and training steps of 0.95 and 1000000, respectively.

**MD simulation**

The LAMMPS software package was used to perform MD simulations. For EMD simulations, the structure was relaxed at canonical (NVT) ensembles with periodic boundary conditions for 0.1 ns with a time step of 1 fs. Then, the heat flux data can be started to collect. For NEMD simulations, first, the structure relaxed at isothermal–isobaric (NPT) ensemble. Then, the atomics of the two sides of the simulation system were fixed and 22 slabs were adopted to the remaining region. Heat source and heat sink is set to 320 and 280 K, respectively. For the further relaxation with canonical (NVT) ensembles, the relaxation time is set as 1 ns. Finally, the temperature of each slab and the heat flux across the system can be calculated.

**FEM simulation**

To construct polycrystalline models, ABAQUS/Standard with python script is used. The models are comprised of 2197 individual grains with Voronoi cells feature, which are randomly assigned with BAs or GaN attributes. Besides, the ITR between BAs and GaN grains are defined based on the NEMD results. Assuming an equivalent average grain with cubic geometry[65], we define the grain size as $\sqrt[3]{\frac{L^3}{N}}$, where the $N$ is the number of grain, and the $L$ is side length. Considering the GaN with anisotropic properties, we assign corresponding grain with random orientations. We apply two highly conductive sheets to construct heat source and cold source. For the initial load condition, we set the temperature of cold source as zero.

# Acknowledgement

We acknowledge the support of the National Natural Science Foundation of China (Grant Nos. 52006057), National Key R&D Program of China (2023YFB2408100), the Fundamental Research Funds for the Central Universities (Grant No. 531119200237), the Natural Science Foundation of Chongqing, China (No. CSTB2022NSCQ-MSX0332), and the State Key Laboratory of Advanced Design and Manufacturing for Vehicle Body at Hunan University (Grant No. 52175013). This work was also supported in part by the NSF (award number 2110033). The numerical calculations in this paper have been done on the supercomputing system of the National Supercomputing Center in Changsha.

# Data availability

The datasets (NNP models, LAMMPS input files, ShengBTE output files and FE models) in this study are available at https://github.com/WU2JING/H.

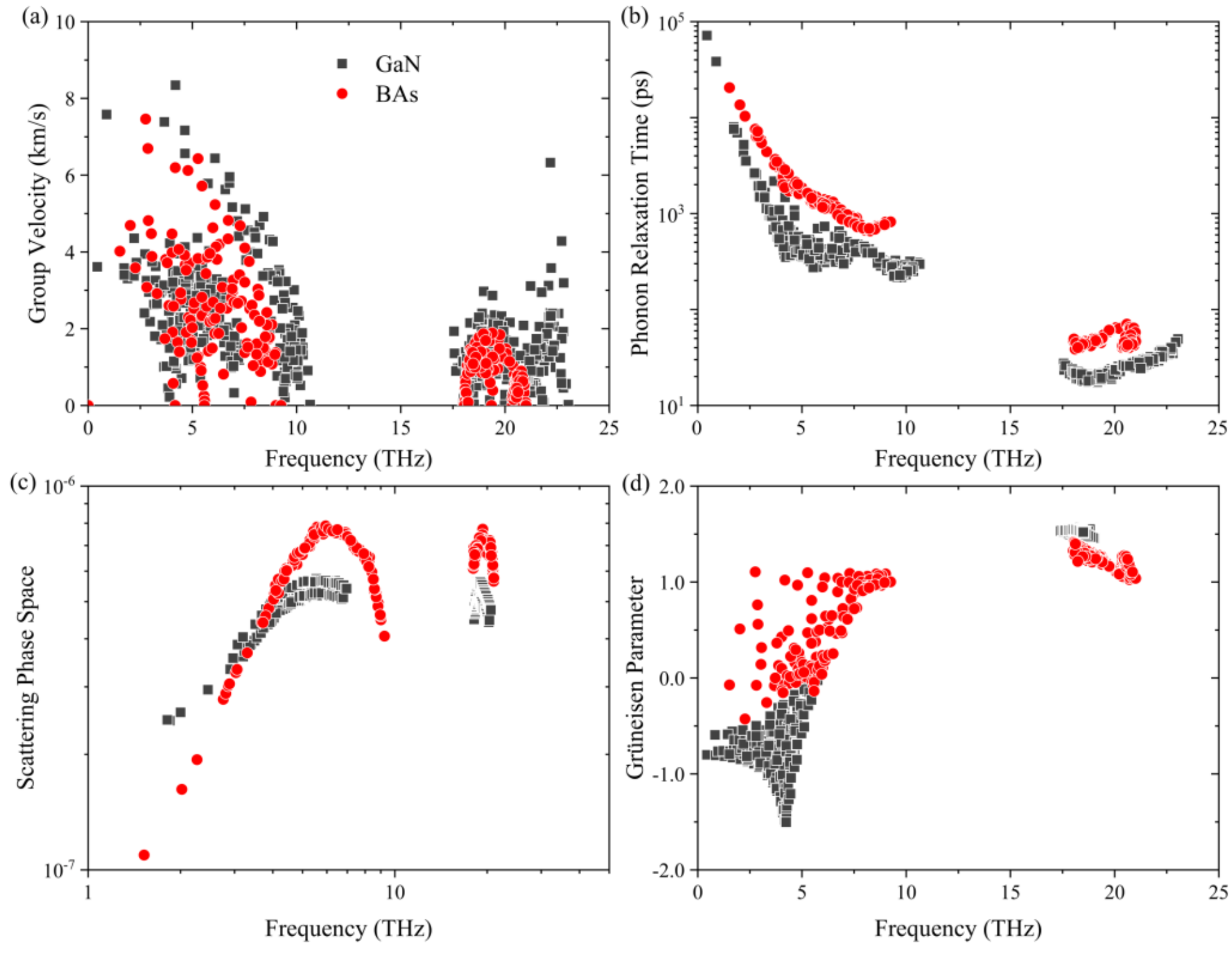

**Fig. S1** The comparison of group velocity, phonon relaxation time, scattering phase space, and Grüneisen parameter between BAs and GaN.

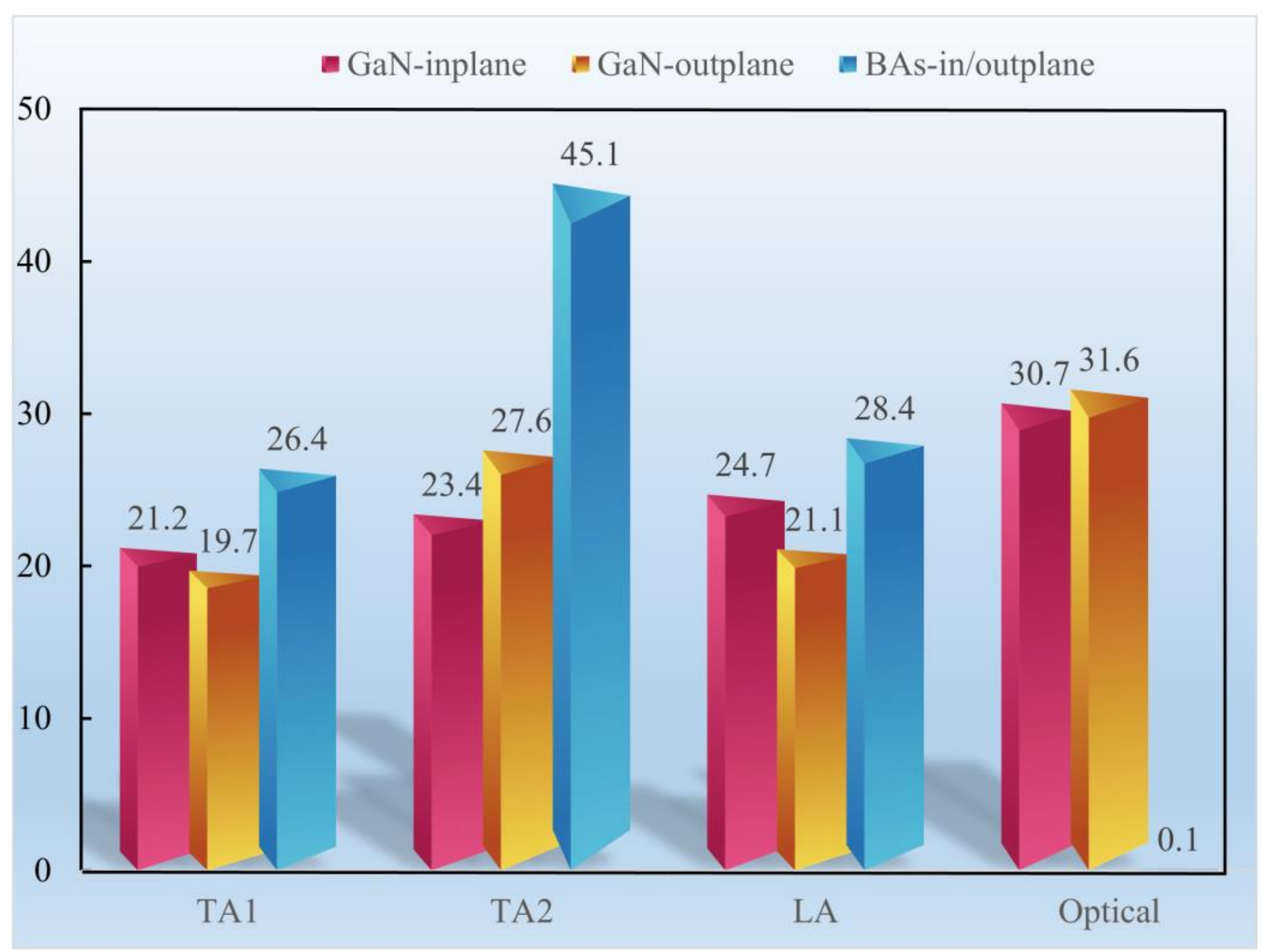

**Fig. S2** The percentage contribution of $\kappa$ from each phonon branch (TA1, TA2, LA and optical).

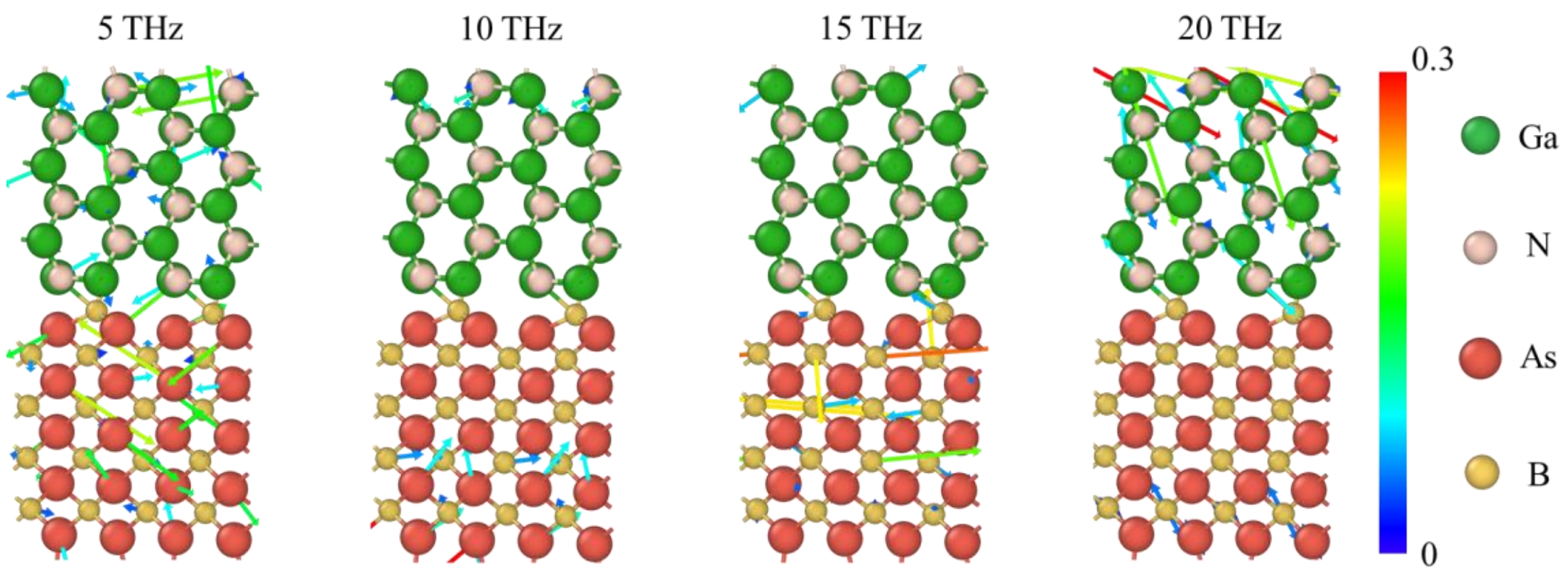

**Fig. S3** The visualization of phonon eigenvectors for GaN-BAs interface at different frequencies.

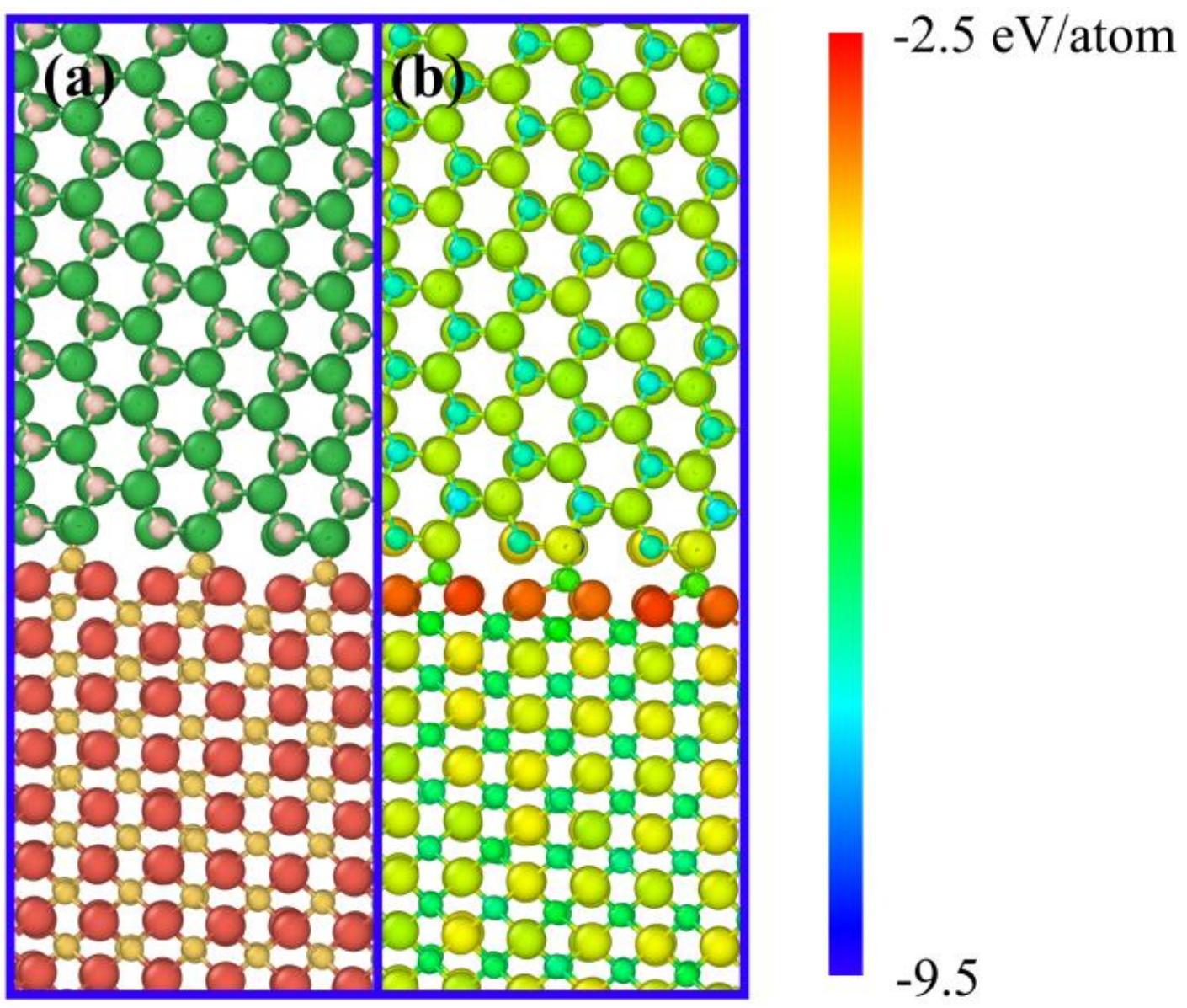

**Fig. S4** (a) GaN-BAs heterostructures after relaxation and (b) the energy distribution.

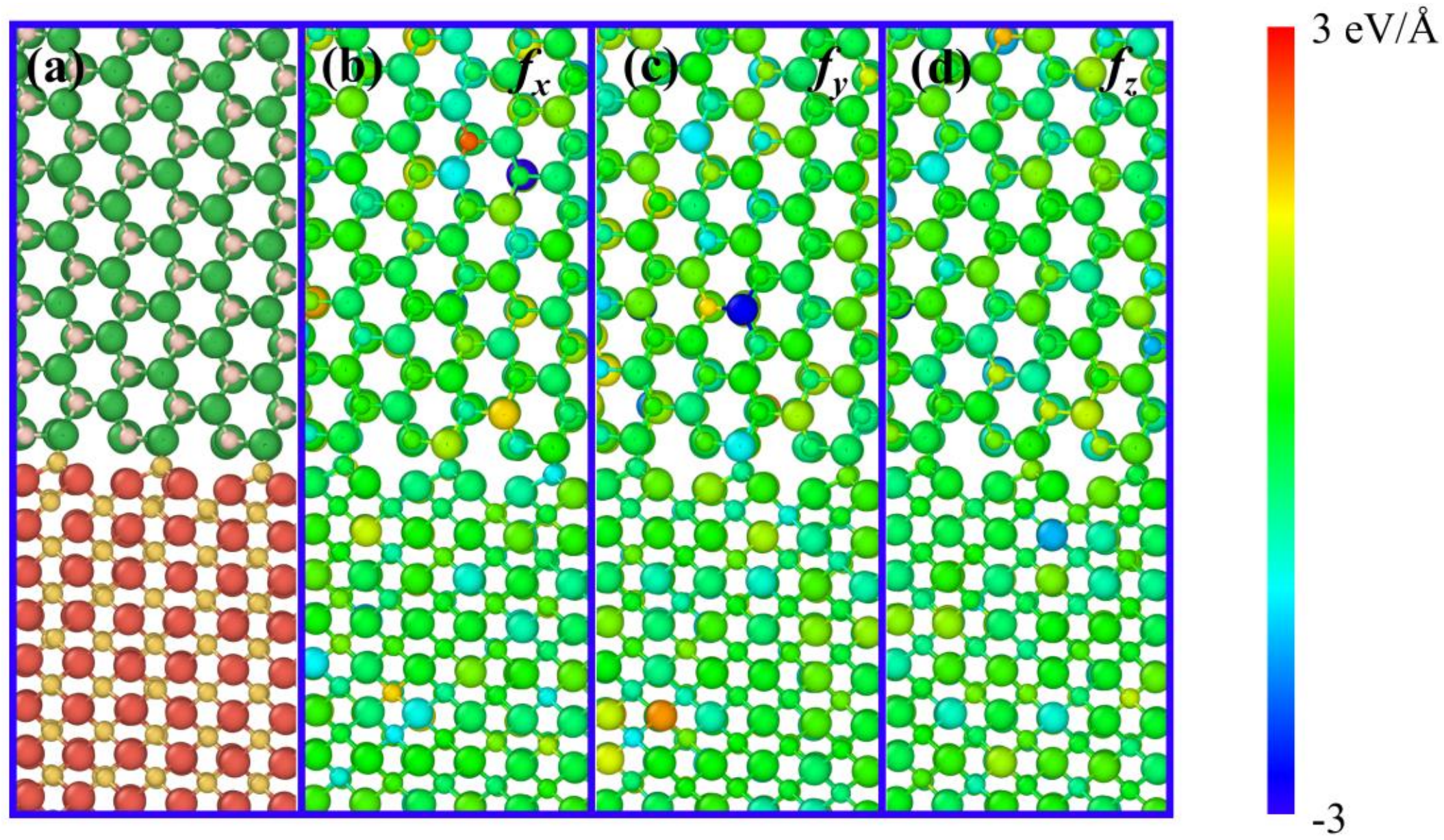

**Fig. S5** Force distribution in GaN-BAs heterostructures. (a) GaN-BAs heterostructure after relaxation. (b) Stress $f_x$ in the $x$-direction of GaN-BAs; (c) Stress $f_y$ in the $y$-direction of GaN-BAs; (d) Stress $f_z$ in the $z$-direction of GaN-BAs.

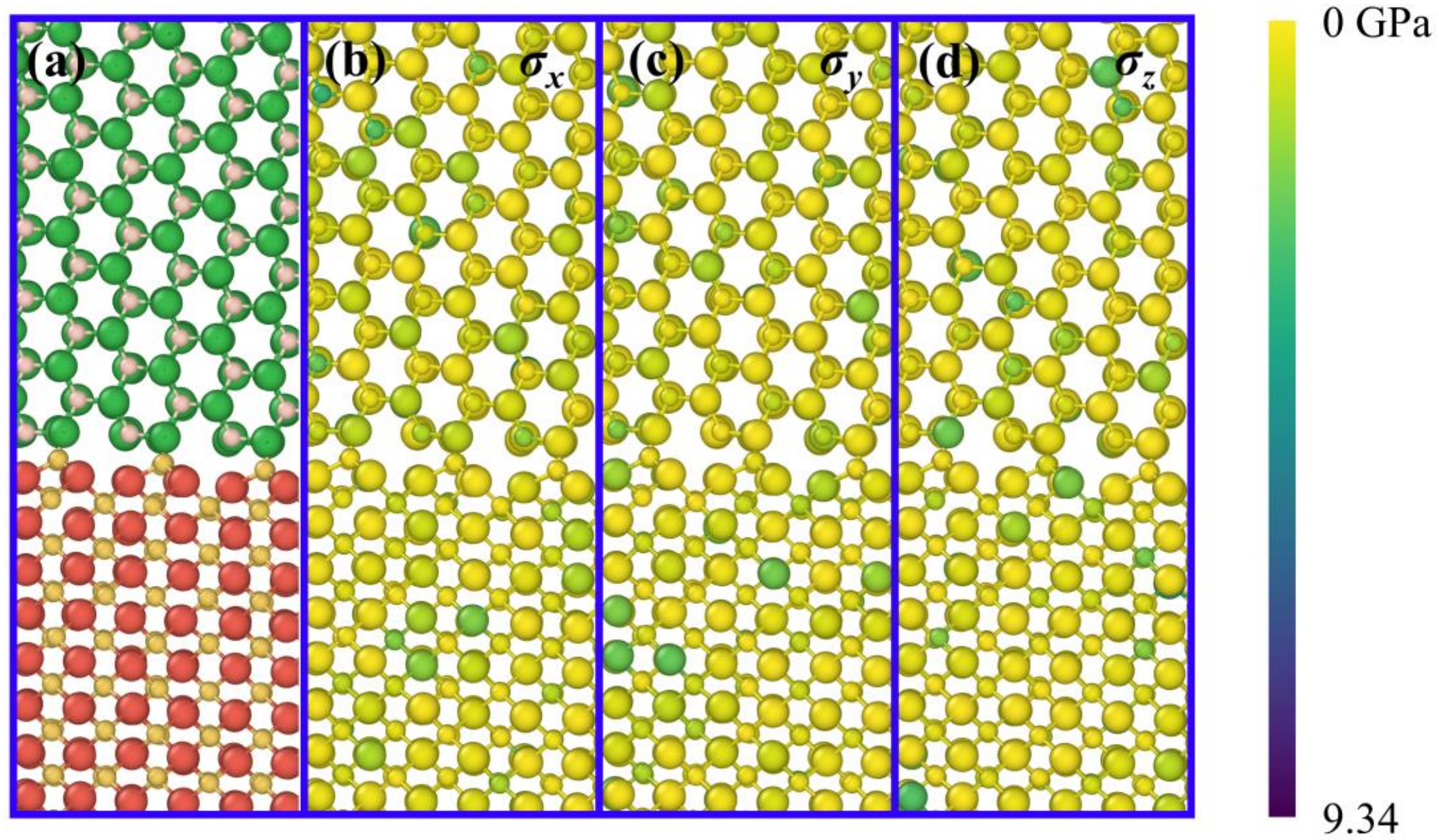

**Fig. S6** Stress distribution in GaN-BAs heterostructures. (a) GaN-BAs heterostructure after relaxation. (b) Stress $\sigma_x$ in the *x*-direction of GaN-BAs; (c) Stress $\sigma_y$ in the *y*-direction of GaN-BAs; (d) Stress $\sigma_z$ in the *z*-direction of GaN-BAs.

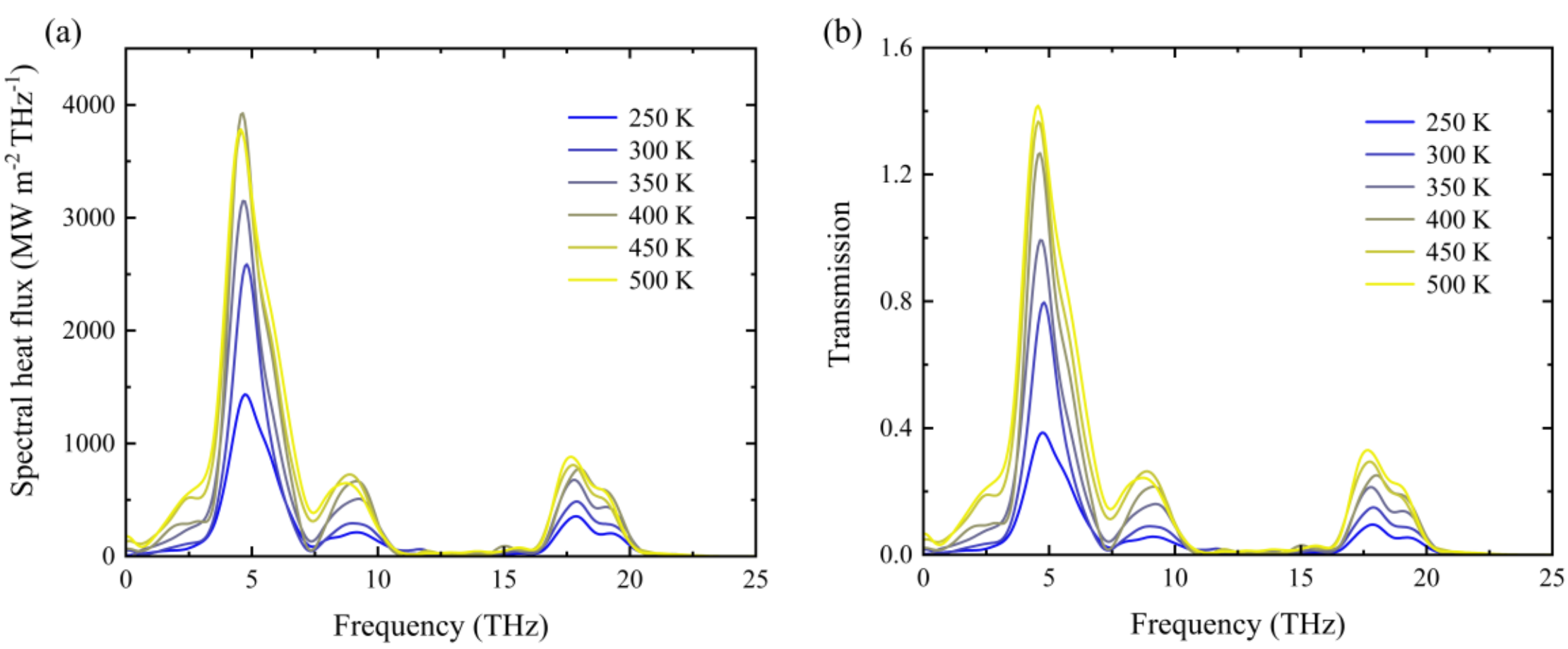

**Fig. S7** The temperature-dependent spectral heat flux and phonon transmission coefficient in GaN-BAs heterostructures.

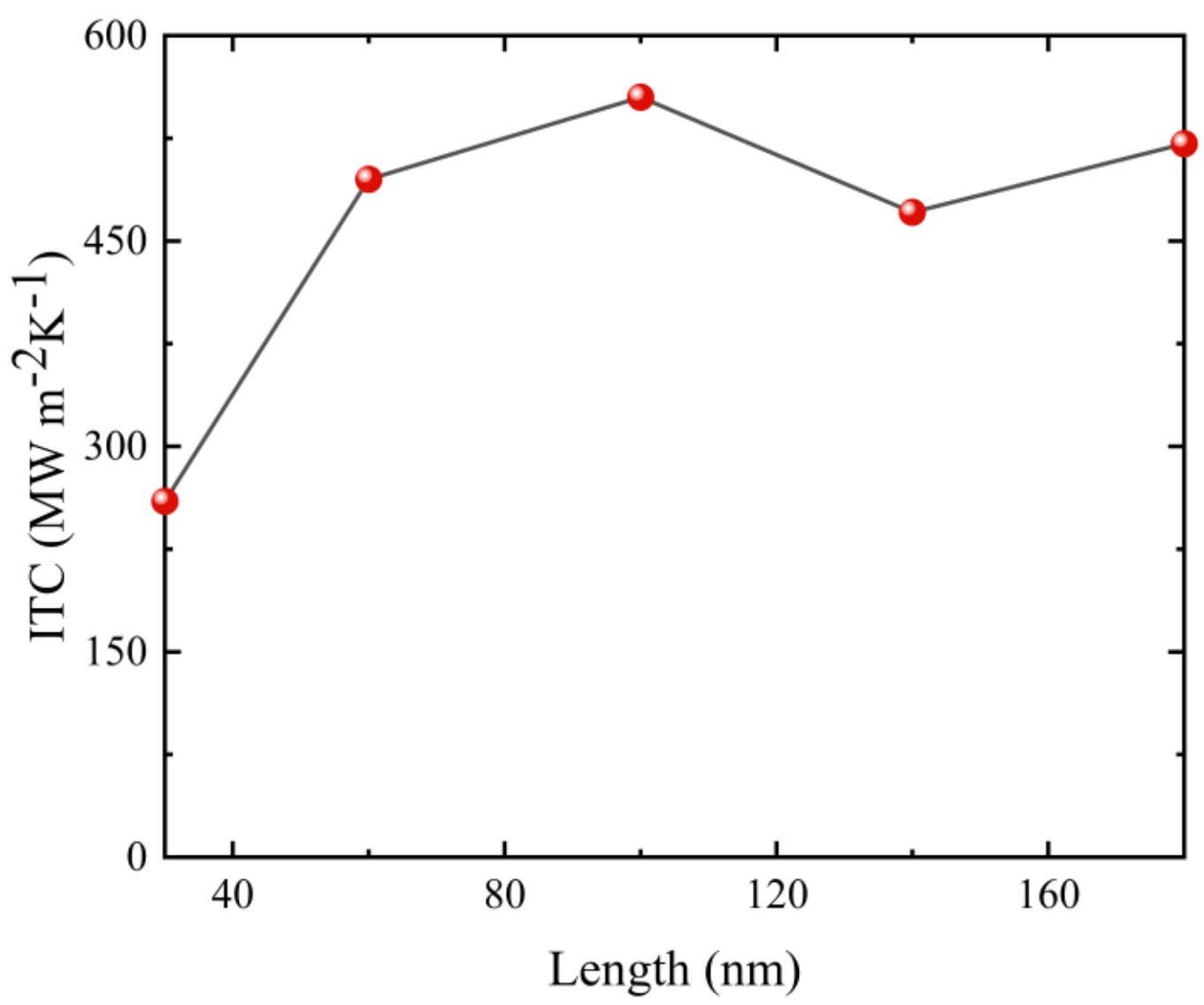

**Fig. S8** The length-dependent ITC in GaN-BAs heterostructures.

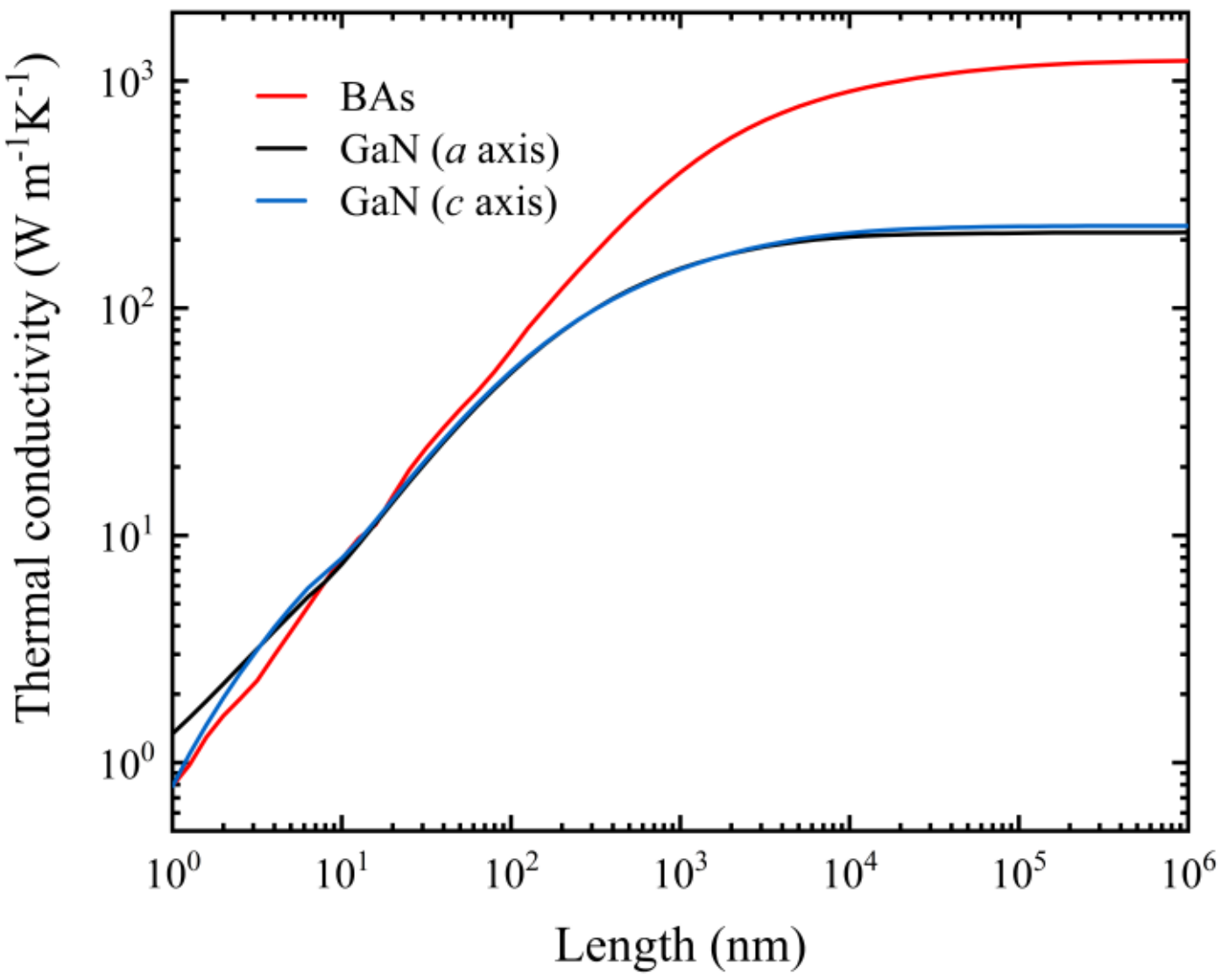

**Fig. S9** The length-dependent thermal conductivity of GaN and BAs.